\documentstyle[]{mn}
\newif\ifAMStwofonts

\ifoldfss
  \ifCUPmtlplainloaded \else
    \NewTextAlphabet{textbfit} {cmbxti10} {}
    \NewTextAlphabet{textbfss} {cmssbx10} {}
    \NewMathAlphabet{mathbfit} {cmbxti10} {} 
    \NewMathAlphabet{mathbfss} {cmssbx10} {} 
  \fi
  \ifAMStwofonts
    \ifCUPmtlplainloaded \else
      \NewSymbolFont{upmath} {eurm10}
      \NewSymbolFont{AMSa} {msam10}
      \NewMathSymbol{\upi}     {0}{upmath}{19}
      \NewMathSymbol{\umu}     {0}{upmath}{16}
      \NewMathSymbol{\upartial}{0}{upmath}{40}
      \NewMathSymbol{\leqslant}{3}{AMSa}{36}
      \NewMathSymbol{\geqslant}{3}{AMSa}{3E}

    \fi
  \fi
\fi 

\ifnfssone
  \newmathalphabet{\mathit}
  \addtoversion{normal}{\mathit}{cmr}{m}{it}
  \addtoversion{bold}{\mathit}{cmr}{bx}{it}
  \newmathalphabet{\mathbfit} 
  \addtoversion{normal}{\mathbfit}{cmr}{bx}{it}
  \addtoversion{bold}{\mathbfit}{cmr}{bx}{it}
  \newmathalphabet{\mathbfss} 
  \addtoversion{normal}{\mathbfss}{cmss}{bx}{n}
  \addtoversion{bold}{\mathbfss}{cmss}{bx}{n}
  \ifAMStwofonts
    \ifCUPmtlplainloaded \else
      \UseAMStwoboldmath
      \makeatletter
      \new@mathgroup\upmath@group
      \define@mathgroup\mv@normal\upmath@group{eur}{m}{n}
      \define@mathgroup\mv@bold\upmath@group{eur}{b}{n}
      \edef\UPM{\hexnumber\upmath@group}
      \new@mathgroup\amsa@group
      \define@mathgroup\mv@normal\amsa@group{msa}{m}{n}
      \define@mathgroup\mv@bold\amsa@group{msa}{m}{n}
      \edef\AMSa{\hexnumber\amsa@group}
      \makeatother
      \mathchardef\upi="0\UPM19
      \mathchardef\umu="0\UPM16
      \mathchardef\upartial="0\UPM40
      \mathchardef\leqslant="3\AMSa36
      \mathchardef\geqslant="3\AMSa3E
    \fi
  \fi
\fi 

\ifnfsstwo
  \DeclareMathAlphabet{\mathbfit}{OT1}{cmr}{bx}{it}
  \SetMathAlphabet\mathbfit{bold}{OT1}{cmr}{bx}{it}
  \DeclareMathAlphabet{\mathbfss}{OT1}{cmss}{bx}{n}
  \SetMathAlphabet\mathbfss{bold}{OT1}{cmss}{bx}{n}
  \ifAMStwofonts
    \ifCUPmtlplainloaded \else
      \DeclareSymbolFont{UPM}{U}{eur}{m}{n}
      \SetSymbolFont{UPM}{bold}{U}{eur}{b}{n}
      \DeclareSymbolFont{AMSa}{U}{msa}{m}{n}
      \DeclareMathSymbol{\upi}{0}{UPM}{"19}
      \DeclareMathSymbol{\umu}{0}{UPM}{"16}
      \DeclareMathSymbol{\upartial}{0}{UPM}{"40}
      \DeclareMathSymbol{\leqslant}{3}{AMSa}{"36}
      \DeclareMathSymbol{\geqslant}{3}{AMSa}{"3E}
    \fi
  \fi
\fi 

\ifCUPmtlplainloaded \else
  \ifAMStwofonts \else 
    \def\upi{\pi}
    \def\umu{\mu}
    \def\upartial{\partial}
  \fi
\fi

\title [Triggered Star Formation in LMC4]
{Triggered Star Formation in the LMC4/Constellation III
Region of the Large Magellanic Cloud}

\author[Y. Efremov and B. Elmegreen]
  {Yuri~N.~Efremov$^1$ and Bruce~G.~Elmegreen$^2$\\
  $^1$ MSU, P.K.Sternberg Astronomical Institute, Moscow 119899\\
  $^2$ IBM Research Division, T.J. Watson Research Center,
        P.O. Box 218, Yorktown Heights, NY 10598}

\date{Accepted 24 April 1998.
      Received 3 March 1998;
      in original form 27 June 1997}

\pagerange{\pageref{firstpage}--\pageref{lastpage}}
\pubyear{1998}

\begin{document}

\maketitle

\label{firstpage}

\begin{abstract}
The origin of a regular, 600 pc-long arc of young stars and clusters in
the Constellation III region of the Large Magellanic Cloud is
considered. The circular form of this arc suggests that the prestellar
gas was uniformly swept up by a central source of pressure. In the
center of the arc are six $\sim30$ My old A-type supergiant stars and
a Cepheid variable of similar age,
which may be related to the source of this pressure. We
calculate the expansion of a bubble around a cluster of this age, 
and show that
it could have triggered the formation of the arc at the right time and
place. Surrounding the central old stars and extending well outside
the young arc is the LMC4 superbubble and giant HI shell. We show how
this superbubble and shell could have formed by the continued expansion
of the 15 My old cavity, following star formation in the arc and the
associated new pressures. The age sequence proposed here was not evident
in the recent observations by Olsen et al. and Braun et al. because the
first generation stars in the center of the LMC superbubble are
relatively faint and scarce
compared to the more substantial population of stars
less than 15 My old that formed throughout the region in a second
generation. These considerations lead to an examination of the origin of
the LMC4/Constellation III region and other large rings in the LMC and
other galaxies. Their size and circularity could be the result of low 
galactic shear and a thick disk, with several 
generations of star formation
in their interiors now too faint to see. 
\end{abstract}


\section{Introduction}

Two large arcs of young stars and clusters are prominently situated in
the northeast corner of the Large Magellanic Cloud (LMC) (see Fig. 1).
One is large and thick, being a quarter
segment of a ring (designated the "Quadrant" here),
and usually identified with Shapley's Constellation III.
The other, slightly to the southwest, is smaller and contains brighter
clusters shaped like one-sixth part of a ring ("Sextant").
The Quadrant consists of Lucke \& Hodge (1970)
associations LH 65, 77 and 84 and is often referred to as LH 77; the
Sextant includes LH 51, 54, 60 and 63 and is not generally considered
noteworthy, being often masked by bright HII regions (cf.  Table 1).
The main peculiarity of both arcs is their very regular circular form.

The region surrounding these arcs is well studied as a possible example
of propagating star formation. Westerlund \& Mathewson (1966) noted on a
UV plate "the great arc of the bright blue stars," saying that "Shapley
called this arc Constellation III"; they also suggested this arc is
connected with the supergiant HI shell that surrounds the most prominent
HI void in the LMC, noted by McGee \& Milton (1966). The inner extent of
this shell, as determined by Kim et al.
(1997), is shown in figure 1 by a circle. 

\begin{figure*}
\vspace {7.5in}
\includegraphics{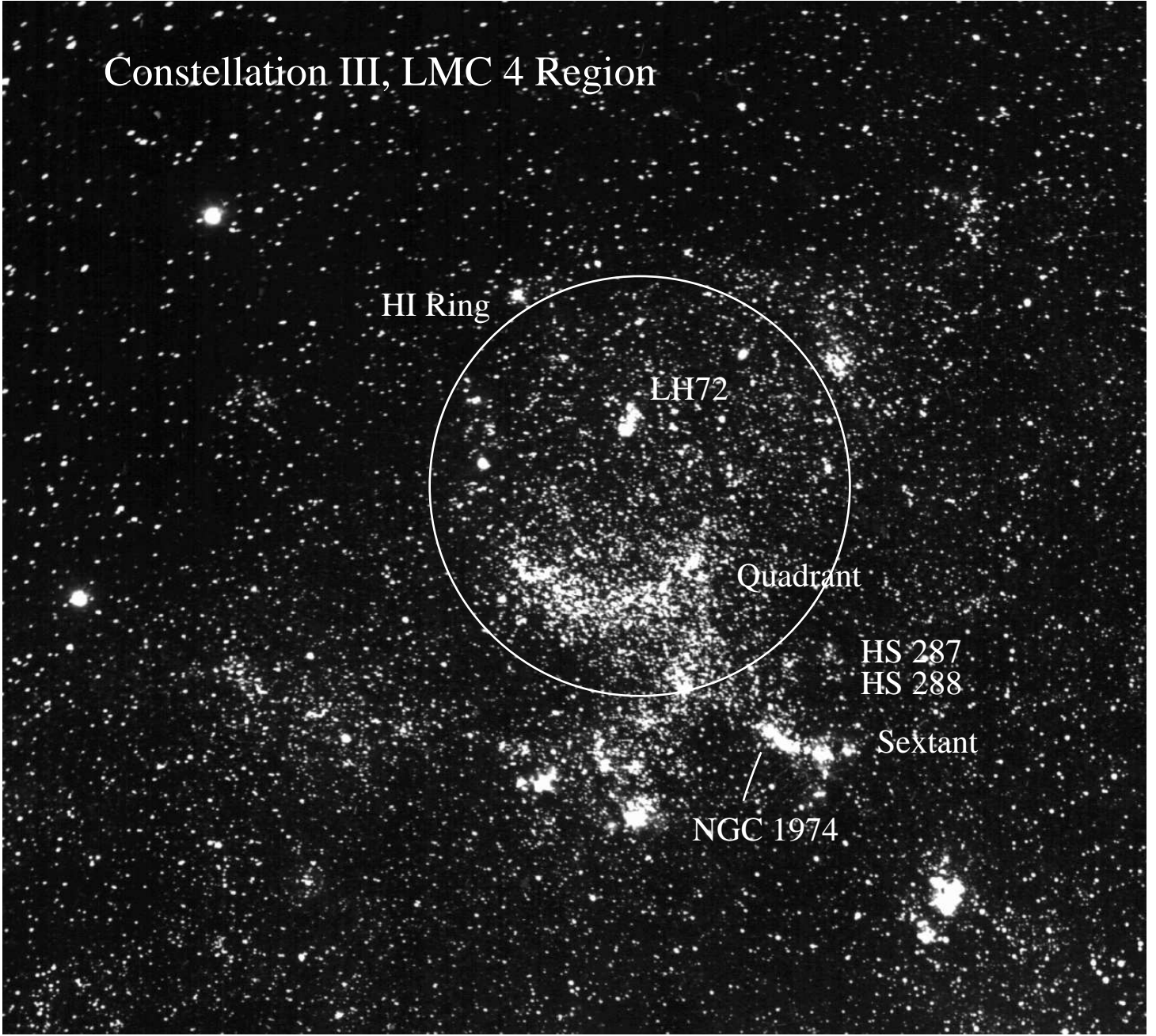}
\caption{Photographic image of the Constellation III region of the LMC with a
circle 
indicating the inner radius of the HI shell and other identifications discussed
in the text. 
The diameter of the circle is 1.32 degrees.
The original photograph is from Boyden Observatory, courtesy of 
Harlow Shapley, Harvard College Observatory, and
kindly 
provided in tiff form by Knut Olsen, Paul Hodge, and Don Brownlee. 
}
\end{figure*}

Westerlund \& Mathewson further suggested that "the annular shell of
neutral hydrogen and stars in the region of Constellation III" is the
remnant of a "super-supernova" outburst (following  Shklovsky 1960).  
They believed that gas was
swept up by the expanding shell; the part of this shell that was
moving to the center of the LMC would accumulate matter more quickly
and be the first to become stationary.	Stars would condense out,
forming the arc of stars of Constellation III.	They believed also
that the southern part of the HI supershell is missing and that "this
gap is filled by the arc of bright blue stars forming Constellation
III."  This latter suggestion is not confirmed by the HI data, which 
shows the southern part of the HI shell clearly
(Domg\"orgen et al. 1995; Kim et al.  1997).

The enormous elongated ring of HII regions around the HI void,
designated superbubble LMC 4, was found later by Meaburn (1980). The HI
void, the surrounding HI shell, and the HII regions and clusters of the
superbubble LMC 4, were all considered by Dopita et al. (1985) to be the
result of supernovae and O-type stars near the center of LMC 4. They
believed they saw an age gradient and determined the velocity of the
star formation front, propagating outward from the center, to be 36
km/s, which was the same as the velocity of the HI shell that they
measured. Subsequent investigations have not confirmed the age
progression or HI velocity (Reid et al. 1987; Domg\"orgen et al. 1995;
Olsen et al. 1997; Braun et al. 1997), though many still believe that
LMC 4 and the ring of young clusters is connected with triggered star
formation. Olsen et al. (1997)
found that the
age of LH72, which is near the center of Constellation III, is about the
same as the ages of the young clusters along the LMC4 ring, although the
age spread in LH72 is large, with stars ranging between $5\times10^6$
and $1.5\times 10^7$ years old. Olsen et al. concluded that their data
are consistent with the alternate model by Reid et al. (1985), in which
star formation proceeded locally for a long time after an initial
trigger from the LMC4 superbubble shock.
Braun et al. (1997) found no signs of an age gradient in the region,
obtaining ages of field stars within a J-shaped strip
inside LMC4, including Quadrant but not LH72.

Domg\"orgen et al.  (1995) studied the origin of the superbubble.  They
considered stellar winds and supernovae in the central association,
the collision of a high velocity cloud with the LMC disk, and the
large-scale propagation of star formation, concluding that the latter
process is more probable.

There is a third, larger, arc to the south of Quadrant, open to the
northeast and touching Quadrant at the northwest. 
It is rather dispersed, and the age spread along
it is large, but at the center of curvature of this arc there is a
massive cluster, NGC 2041, 
that is older than most of the clusters
within the arc. Members of this arc are
listed in Table 1. It may be another example of triggered
star formation, but it is less well defined than Quadrant and
Sextant, so we do not discuss it further.
All three arcs were noted and sketched by Hodge (1967), who also
found similar features in NGC 6946.

Here we consider the possibility that the Quadrant and Sextant arcs
formed by instabilities in a swept-up ring and shell, respectively. We
also show how the giant HI ring surrounding Quadrant could have been
rejuvenated by new pressures from these stars after the first generation
pressures subsided. We then discuss why the LMC4 region appears so
unusual compared to shells and rings in other galaxies. 
We do not comment on the origin of
the first generation of stars (e.g., see de Boer et al. 1998), but only point
out some likely members based on catalogs of supergiants and Cepheid variables.

\section{Star cluster arcs}

We first consider the nomenclature of features in this region.
Shapley's original name of Constellation III was
given to another region, not to what is commonly considered to be
Constellation III.  McKibben Nail \& Shapley (1953) designated
NGC 1974 as the identifier of Constellation III, including an area of 
28' x 28' around NGC 1974
(see Fig. 1). 
They also noted that Constellation III is a
triple cluster, so in fact they were probably referring to Sextant.
This is sensible because Sextant is brighter than Quadrant (cf.
Fig. 1), and would have been more noticeable to McKibben Nail \&
Shapley. Map V45 in the Hodge \& Wright (1967) Atlas of the
LMC clearly shows that NGC 1974 is within Sextant,
and what is commonly called Constellation III, which is the large arc
called Quadrant by us, is not within the 28' x 28' field around NGC
1974.

Westerlund \& Mathewson (1966) were evidently the first to identify
"the arc of bright blue stars" (i.e., Quadrant) with Shapley's
Constellation III.  
Olsen et al.  (1997) also called Quadrant
Constellation III in their figure 1, yet in the text they
refer to the entire superbubble LMC4 as Constellation III.
van den Bergh (1981) called the whole LMC4 region 
Constellation III as well.
This identification was typically the case in papers concerning the
whole region, and was probably one of the reasons why neither Quadrant
nor Sextant have been considered as peculiar features deserving 
study by themselves.

The Quadrant and Sextant arcs are indeed unique features; there is
nothing similar in the LMC.  Quadrant consists of both clusters
and individual stars and may also be designated as LH77 (the small
associations LH84 and LH65 are inside of it).  Sextant includes LH51 =
SL 456, LH54 = NGC 1955, LH60 = NGC 1968 and LH 63 = NGC 1974 (Lucke
\& Hodge 1970, Fig.  1 and Table 1;  NGC 1974 is misprinted as NGC
1947 in this Table).

Similar arcs of star clusters have not been reported elsewhere either.
Besides the NGC 6946 features (Hodge 1967), the only thing resembling
the Quadrant and Sextant stellar arcs is a large
region in the galaxy NGC 1620 studied by Vader \& Chaboyer (1995).  At
the inclination of this galaxy, it is difficult to tell if this is a
triggered arc or a spiral arm.

Data on all three arcs are given in Table 1.  The integral UBV and
position data were taken from Bica et al.  (1996), and the ages were
determined from the U-B and B-V data via the $S$ values according to
Girardi et al.	(1995).  Girardi et al.  introduced $S$ values as a
combination of U-B and B-V integral colors of clusters and calibrated
these values as a function of age, using 24 rich clusters with ages
determined from color-magnitude diagrams.  These clusters gave the $S
- \log t$ relation with an rms dispersion of 0.137 in $\log t$
(Girardi et al.  1995).  For less populous clusters, the error should
be larger.

Only integral colors are available for the bulk of clusters considered
here. After this paper was submitted, the preprint by Braun et al.
(1997) became available. It contains age determinations from CMDs of
several fields inside LMC 4. Fields 0 - 10 of Braun et al. (1997) are
within the Quadrant, and their age range is 9 - 16 Myrs (most are within
the range 10 - 14 Myrs). The $S$ values for these clusters give an age
range within Quadrant (table 1) that is similar, 7 - 21 Myrs (most are
within 10 - 18 Myrs). More detailed comparisons for the Quadrant region
are impossible, because Braun's et al. fields are much larger than the
clusters measured by Bica et al. (1996). For the Sextant region
there are ages obtained by Petr (1994), as given by Braun et al. 1997
(in Myrs): LH63 - 14, LH60 - 9, and LH54 - 6, whereas our values (Table
1) are respectively 7, 3-4 and 3 Myrs. Braun et al. (1997) comment that
the accuracy of their ages is $\sim0.1$ in logarithmic units and because
of the different isochrones used, their ages should be slightly older
(by 0.05 in $\log t$) than those determined by Girardi et al. (1995). We
conclude that ages derived from the integral photometry are consistent
with those from the CMDs and are accurate enough for our purposes.

A remarkable feature of the two arcs is their circularity. This regular
form suggests they were produced from nearly uniform gas swept up by a
central source of pressure. The smaller size and younger age of
Sextant, and its position near the edge of LMC4, suggest that it was
triggered inside the dense HI shell that was swept up earlier to make
Quadrant.

\section{Possible pressure sources}
\label{sec:pressource}

The radius of the Sextant arc is $\sim170$ pc and at its center is the
small cluster HS288 (cf. Fig. 1). 
At 70 pc to the north there is a larger dispersed
cluster HS287 surrounded by the HII region N50. For the Quadrant arc the
radius is $\sim280$ pc and there is nothing obvious at the exact center,
but at $\sim160$ pc to NNW from the center of curvature there is a bright
cluster LH72 with an age range of 5 - 15 Myrs (Olsen et al. 1997)
surrounded by the HII region N55.  
These sizes assume a distance to the 
LMC of 45 kpc (Berdnikov, Vozyakova, \& Dambis 1996;
Efremov 1997; Efremov, Schilbach, \& Zinnecker 1997; Fernley et al. 1998).

The radius of the Quadrant arc is too large, and the projected expansion
speed of the HI shell too small ($<10$ km s$^{-1}$; Domg\"orgen et al.  1995),
for a young cluster like LH72 to have formed it. 
Besides, the LH72 cloud in Kim et al. (1997) 
looks like a bright rim that
is part of a shell
coming off from the northeast rim of the LMC4 superbubble, so it is not
related to the trigger for Quadrant or the expansion of LMC4.

Sextant is somewhat different: the arc is
small and the local density is high, so the formation time could have
been much shorter than for Quadrant. 
There are no age data for the two clusters HS 287 and HS 288 near the
center of Sextant, but HS 287 must contain at least some young stars,
with ages of $10^7$ years or less, considering the ionization in the
associated nebula N50.	
There is also ionization in Sextant, forming a
shell around its eastern end, which is N51D (Meaburn \& Terret 1980).
This is consistent with both its younger age and the higher density of
ambient HI compared to Quadrant.

Thus we consider the possibility
that the clusters HS287 and/or HS288 formed the Sextant arc in only 10
My or so, and that older, fainter stars near the center of curvature of
the Quadrant arc formed this larger region. 

There are indeed
old stars near the center of the Quadrant arc. At the
center of the LMC4 supershell there is a sparse faint cluster HS 343,
for which no photometry exists. HS 343 could be much
older than LMC4, in which case it would be irrelevant here. 
(HS 343 is not visible in Fig. 1, so its position is indicated in Fig. 2.)

More important is 
a small grouping of 6
A-type supergiant stars in the catalogue of Rousseau et al. (1978)
within a circle of $\sim10$ arcmin diameter near the center
of curvature of Quadrant. These stars have
apparent V magnitudes of 12.1 to 12.5, as given in Table 2. 
This is the only concentration of
type AI or later supergiants within the LMC4 area, as is evident from figure 2,
which shows all of the A-type supergiants in the region. 
There are also
several B and M-type supergiants near the center of LMC4, from the
catalogs of Rousseau et al. (1978) and Rebeirot et al. (1983), respectively, but
these stars show no particular concentration like the A stars. 

\begin{figure}
\vspace {3.0in}
\includegraphics{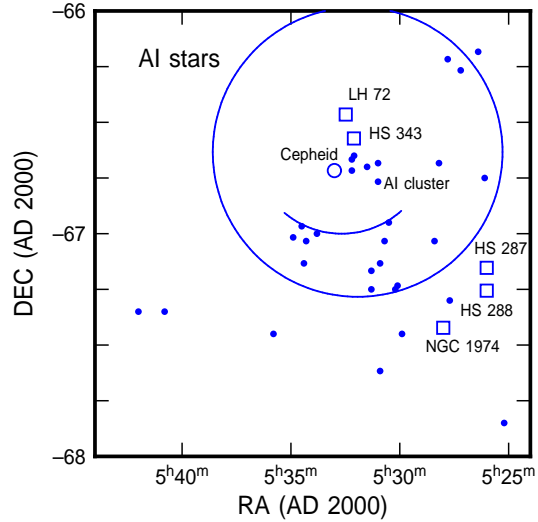}
\caption{A-type Supergiant stars in the vicinity of Constellation III,
from the catalog of Rousseau et al. (1978).
The open circle is the Cepheid variable star HV5924, and the open
squares are some of the clusters discussed in the text.
}
\end{figure}

Isochrones by Bertelli et al. (the same as those used for the cluster
ages here and in Girardi et al.) give $\log t=7.4$ for AI stars, 7.5 for
BI stars and about 7.6 (V magnitudes are uncertain) for MI stars. The
positions of these LMC4 supergiants in a color magnitude diagram are
very similar to the positions of the brightest stars in two LMC young
globulars, NGC 2004 and NGC 2100. These clusters were used by Girardi et
al. to calibrate integral colors as ages. The average ages of the
clusters from their CMDs are $\log t= 7.33$ and 7.40,
whereas their ages
from the blue supergiants are $\log t=7.66$ and 7.71. The difference is
explained by much younger ages for the main sequence turn-off points,
6.99 and 7.10 respectively. This difference illustrates the age range
for star formation, and the uncertainties present in the theory of 
supergiants.
To be more in accordance with the ages from the integral
colors, we prefer the ages that are connected with the supergiants
stars, because these contribute most strongly to the integral colors of
a cluster. This age is $\sim30$ My, and clearly larger, by $\sim15$
Myrs, than the ages of most of the Quadrant stars.

A Cepheid star, HV5924, with a period 16.2 days, is close to 
HS343 and
the AI supergiants; its position is shown in Figure 2 by an open circle. 
This period corresponds to an age of 75 My, according to a recent
calibration in Efremov \& Elmegreen (1998). Two other Cepheids, HV 5921
and HV 12436, are also nearby, but their periods are much shorter, 3.0
and 4.3 days, and so their ages larger, 140-180 My. Considering the
inaccuracy in Cepheid ages, HV5924 could be coeval with the AI stars,
but the other two Cepheids are probably not. 

In view of the presence of luminous stars near the center of LMC4,
there must have been significant star-forming activity there
about 30 My ago. This is presumably the
event whose pressures led to the formation of the Quadrant arc, and
which also began the current generation of star formation all throughout
this region.
The cluster of 6 AI stars is not at the exact center of the Quadrant 
arc, but this could be the result of a small
southernly drift of this cluster
from a $\sim4$ km s$^{-1}$ motion of its primordial cloud.

\section{Triggered star formation in Quadrant}

\subsection{Triggering Conditions}

We propose that the Quadrant arc of young stellar clusters was formed by
the gravitational collapse of swept-up gas in the densest part of an
expanding {\it ring} that surrounded what was then a $\sim14$ My old
cluster, but which is visible today as a dispersed group of supergiants
and Cepheid variables, $\sim30$ My years old. 
The current age of the Quadrant ring is taken to be 16 My, from the
oldest stars in this region found by Braun et al. (1997).
The total age of $\sim30$ My is imprecise. The solutions
given below 
for the triggering and expansion of the Quadrant arc and HI ring depend
on this age, but the qualitative
nature of the results, i.e., our proposal that stars connected with 
the cluster of AI supergiants triggered the formation of Quadrant,
does not depend on this total age very much unless it is wrong by more than 
a factor of two. 

The ring geometry assumed here is
important, as opposed to a three-dimensional shell geometry, because the
divergence of gas in a ring differs from the divergence in a shell,
giving a different equation for the collapse time (Elmegreen 1994).
Rings are also more likely than shells for large disturbances in galaxy
disks because the gravity of the disk pulls high latitude shell material
down to the midplane where it makes a ring, and because most of the
ambient gas that is compressed by the disturbance is close to the plane
(Ehlerov\'a et al. 1997).

The formation of giant molecular clouds in expanding {\it shells} of gas
has been considered a mechanism for triggered star formation since
the work of Tenorio-Tagle (1981) and Elmegreen (1982ab). The first
analytical work on gravitational instabilities in expanding shells was
by Ostriker \& Cowie (1981) and Vishniac (1983). A more detailed
analytical analysis of shell expansion and collapse, with applications
to triggered star formation in galaxies, was in McCray \& Kafatos
(1987). The first work on gravitational instabilities in expanding rings
was in Elmegreen (1985). The basic idea in these references is the same
as that applied here to the Quadrant and Sextant arcs.

In this model, gas that is collected into a compressed ridge by a moving
shock front becomes gravitationally unstable because of its high
density, and it forms one or more cores with even higher density. Stars
form in these high density cores by normal processes, but at a much
higher rate than would have occurred without the compression.  This is
the triggering model introduced by Elmegreen \& Lada (1977). It is
distinct from another prominent triggering mechanism in which
pre-existing clouds are imploded by a shock front directly. The
implosion model dates back to Dibai (1958) and Dyson (1968), with recent
work by Lefloch \& Lazareff (1994), Boss (1995), and others.
It was applied to
the Constellation III region by Dopita et al. (1985). The first model,
sometimes called the "collect and collapse" model, is preferred to the
implosion model for the Quadrant region because Quadrant consists
of semi-regularly spaced clusters with a uniform size strung out along an
arc, in addition to many bright field stars.
This is precisely the geometry expected for an instability in a
swept-up piece of a ring; each cluster formed at a local center of collapse.
Implosion models should give a more irregular
placement of second generation regions, as Dopita et al. discussed. A
review of both triggering mechanisms is in Elmegreen (1998).

Here we consider two models for a ring expanding into a uniform medium.
The first has a shock velocity $V\propto t^{-0.4}$ and radius $R\propto
t^{0.6}$ in the case where energy is continuously put into a spherical
cavity from combined supernovae and stellar winds 
(Pikelner 1968; Dyson 1973;
Castor, McCray \&
Weaver 1975; for a more recent discussion, see
Comer\'on, Torra, \& G\'omez 1998). 
The three-dimensional shape of the cavity may not be this
simple, however, because the perpendicular extent may be larger than the
in-plane extent (e.g. Tenorio-Tagle \& Bodenheimer 1988). An alternate
model considers that the energy in a cylinder increases linearly with
time, rather than the energy in a sphere. The mass of a cylinder
increases as $R^2$ so the energy constraint in this case gives
$R^2V^2\propto t$ instead of $R^3V^2\propto t$ for a sphere.	The
similarity solution for the expansion is then $R\propto t^{3/4}$ and
$V\propto t^{-1/4}$. These two cases will be designated as spherical and
cylindrical energy input, respectively. For a general derivation of the
collapse conditions, we write $R\propto t^\kappa$, where $\kappa=3/5$
and $3/4$ in the spherical and cylindrical geometries, respectively.

The collapse time for a ring with an $R\propto t^\kappa$ expansion law
may be derived following the example in
Elmegreen (1994).  The model considers the
accumulation
and divergence of material in a ring with half-thickness $r$,
compressed density $\rho$, mass per unit length
$\mu_0=\rho\pi r^2=\rho_0Rr$,
turbulent velocity dispersion in the ring, $c$, angular velocity of
the growing perturbation, $\Omega$,
and preshock density $\rho_0=n_0m_{\rm
H}$.  With perturbed quantities indicated by a subscript 1, the
equations of motion and continuity for the growth of transverse
perturbations are
\begin{equation}
\mu_0R{{\partial \Omega}
\over{\partial t}}=-c^2 \nabla\mu_1 +\mu_0
g_1 -3\mu_0\Omega V,
\end{equation}
\begin{equation}
{{\partial \mu_1}\over{\partial t}}=-\mu_0 R\nabla\bullet \Omega -
\mu_1{{V}\over{R}}.
\end{equation}
The self-gravitational acceleration $g_1=2G\mu_1 k \ln (2/kr)$ is for
a perturbation along the periphery of the ring with wavenumber
$k=2\pi/\lambda$ and wavelength $\lambda$; this approximation is
reasonable for $2>>kr$.

We consider sinusoidal perturbations and let the time derivatives in
the above equations equal the instantaneous growth rate $\omega$.
Then $\Omega$
and $\mu_1$ can be eliminated and an equation for $\omega$ as
a function of $k$ results (see Elmegreen 1994):
\begin{equation}
\omega=-{{2V}\over{R}}+\left({{V^2}\over
{R^2}}+2G\mu_0k^2\ln(2/kr)-c^2k^2
\right)^{1/2}.\end{equation}
Setting $d\omega/dk=0$ gives the fastest growing wavelength, and
after substituting this into the equation for $\omega$ we
get the fastest growth rate:
\begin{equation}
\omega_{\rm peak}=-{{2V}\over{R}}+
\left({{V^2}\over{R^2}}+G\mu_0k_{\rm peak}^2
\right)^{1/2},\end{equation}
where
\begin{equation}
{{k_{\rm peak}r}
\over{2}}=\exp\left[-0.5\left(1+{{c^2}\over{G\mu_0}}\right)\right].
\label{eq:k}
\end{equation}

Now we set the fastest growth rate equal to the inverse of the ring
age in order to get the time of significant collapse after the
beginning of the expansion; for $R\propto t^\kappa$, this
time is $t=\kappa R/V$ (see also
Theis et al. 1997; Ehlerov\'a et al. 1997).
This leads to an equation for $t$ that has
to be solved iteratively since $\mu_0$ in the exponent of equation
(\ref{eq:k}) depends on $t$, i.e.,
$\mu_0=\rho\pi r^2=\rho_0{\cal M}^2\pi r^2$
for shock speed $V$ a function of $t$; here we have
assumed a shock compression of $\rho/\rho_0=(V/c)^2\equiv{\cal M}^2$.
The result is
\begin{equation}
t \exp \left[-0.5\left(1+{{c^2}\over{G\mu_0}}\right)\right]=
{{([1+2\kappa]^2-\kappa^2)^{1/2}}
\over{(4\pi G\rho_0)^{1/2}{\cal M}}}.\end{equation}

To solve this, we write
\begin{equation}
t={{T_0}\over{(G\rho_0)^{1/2}{\cal M}}},\end{equation}
and introduce the galactic scale height
\begin{equation}
H^2 = {{ c_0^2 (1+\alpha+\beta)}\over { 2 \pi G \rho
_{\rm 0T}}}
\;\;\;{\rm   for}\;\;\;
\alpha = {{ B ^ 2 } \over { 8 \pi P}}\;\;\;{\rm and}\;\;\; \beta =
{{ P _ {\rm CR}} \over P }
\end{equation}
with ambient
magnetic field strength $B$, cosmic ray pressure
$P_{\rm CR}$, turbulent pressure $P$, velocity dispersion
$c_0$, and total gas+star
density $\rho_{\rm 0T}$ in the gas layer.
We also write $\mu_0=R\rho_0H$. With these substitutions,
\begin{equation}
{{c^2}\over{G\mu_0}}={{\kappa(2\pi)^{1/2}c/c_0}\over{T_0
([1+\alpha+\beta]
\rho_0/\rho_{\rm 0T})^{1/2}}}.\end{equation}
Typically, $(1+\alpha+\beta)\rho_0/\rho_{\rm 0T}\sim1$, so
the parameter $T_0$ in the collapse time satisfies the equation
\begin{equation}
T_0 \exp
\left[-0.5\left(1+{{(2\pi)^{1/2}\kappa c}
\over{T_0c_0}}\right)\right]\approx\left({{(1+2\kappa)^2-\kappa^2}\over
{4\pi}}\right)^{1/2} .
\label{eq:t0}
\end{equation}

Equation (\ref{eq:t0}) is solved numerically for $T_0$ as a function
of $c/c_0$.
Then the time for triggering cloud formation in the ring is
\begin{equation} t_{\rm trig}={{T_0}\over{(G\rho_0)^{1/2}{\cal M}}},
\label{eq:t} \end{equation}
and the corresponding radius is
\begin{equation}
R_{\rm trig}={{T_0c}\over{\kappa(G\rho_0)^{1/2}}}.
\end{equation}
In physical units:
\begin{equation}t_{\rm trig}={{80T_0}\over{n_H^{1/2}{\cal
M}}}\;\;\;{\rm Million\; years} \label{eq:t1} \end{equation}
\begin{equation}R_{\rm trig}={{80T_0c }\over{\kappa n_H^{1/2}}}\;\;{\rm pc},
\label{eq:r} \end{equation}
for $T_0$ of order unity, $c$ in km s$^{-1}$,
and ambient hydrogen density $n_H$ in cm$^{-3}$ (considering 
He also in $\rho_0$).
Note that if the gas were in the shape of a sphere instead of a
ring, the time of collapse would be proportional to
${\cal M}^{-1/2}$ instead of
${\cal M}^{-1}$, and the radius of collapse would be proportional to
${\cal M}^{1/2}$.

The age, $t_{\rm now}$, and radius, $R_{\rm now}$, of the Quadrant ring are
currently larger than the triggering
age and radius because the triggering happened some time ago when the
clusters in the arc began forming.
To explain the current radius of the arc, we have to consider how
the stars moved after star formation
began. We consider three cases: First, the arc continued to
expand as $R\propto
t^\kappa$ for the two expansion geometries
discussed above (spherical and cylindrical, respectively),
which means the cavity continued to inflate with
energy from the old clusters and the arcs also continued to accumulate
gas.  This gives
\begin{equation}R_{\rm now}=R_{\rm trig}\left({{t_{\rm now}}\over
{t_{\rm trig}}}\right)^{\kappa}.\label{eq:rnow1}\end{equation}
Alternatively, we
assume that the source energy input stopped when star formation
began and the arcs
decelerated faster with the continued accumulation of gas
(snowplow model). Then momentum conservation gives
$R^Dv=$ constant in the spherical ($D=3$) and
cylindrical ($D=2$)
energy input models, so
\begin{equation}R_{\rm now}^{D+1}=
R_{\rm trig}^{D+1}+(D+1)R_{\rm trig}^DV_{\rm trig}
\left(t_{\rm now}-t_{\rm trig}\right).\label{eq:rnow2}\end{equation}
A third possibility is
that the arcs drifted with constant speed
after they
began forming stars, as if the gas dynamics were not important anymore.
Then
\begin{equation}R_{\rm now}=R_{\rm trig}+V_{\rm trig}
\left(t_{\rm now}-t_{\rm trig}\right).\label{eq:rnow3}\end{equation}

\subsection{Solutions for the formation and motion of Quadrant}

To solve for the shell dynamics,
we need the ambient hydrogen density, $n_H$.  This is obtained from the
HI contours in
Domg\"orgen, Bomans, \& de Boer (1995)
by dividing the average HI column density far outside LMC4, 
$\sim1.2\times10^{21}$ cm$^{-2}$, by
an assumed disk thickness. We choose a thickness equal to 400
pc, which is slightly larger than the disk thickness in the
inner Milky Way because irregular galaxies tend to have
larger thicknesses than spirals (van den Bergh 1988). The result
is $\sim 1$ cm $^{-3}$ for the Quadrant region.

We also need the velocity dispersion in the
ring, $c$. This is not observed, but we assume the ambient
dispersion is $c_0=5$ km s$^{-1}$ and treat the ratio $c/c_o$
as an adjustable parameter. 

The solution for $R(t)$ is obtained by searching for the value of
$c/c_0$ which gives $T_{\rm trig}=14$ My, the approximate time interval
between the first generation of star formation (the A supergiants, $\sim30$ 
My old) and the oldest stars in Quadrant (16 My from Braun et al. 1997). 
The exact value of this time is not
important here; analogous solutions can be obtained for a range of
values. To find $R(t)$, we have to consider both equations (\ref{eq:t1})
and (\ref{eq:r}) for the collapse condition, and equations
(\ref{eq:rnow1}), (\ref{eq:rnow2}), or (\ref{eq:rnow3}) for the
expansion since the time of collapse, using the present Quadrant radius,
$R_{\rm now}\sim280$ pc, and the total elapsed time $t_{\rm now}=30$ My
(again these values are estimates, meant only to illustrate the
plausibility of the basic model until better data are available for
this region). 

\begin{figure}
\vspace {5.2in}
\includegraphics{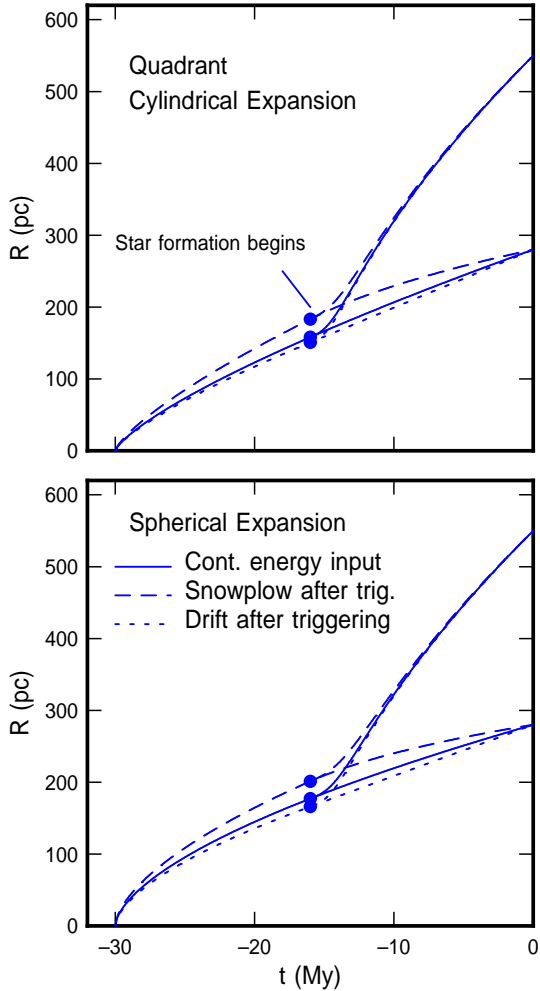}
\caption{The radius of the proposed ring that formed the
Quadrant arc is shown as a function of time for
various models. The dots show the radii and times for
gravitational collapse in the ring, when star formation
in Quadrant is assumed to have begun.
}
\end{figure}

The results for the three expansion cases with spherical and cylindrical
geometries are shown in figure 3. The solutions are very similar because
they are tightly constrained by the collapse and total ages, and by the
present-day radius of Quadrant. The time when Quadrant began forming
stars is indicated by a large dot, $\sim16$ My ago. Time is measured
backwards in this diagram, with the present time equal to 0 and the
events in the past written as negative time. The stellar velocities
today are the slopes of these lines at $t=0$; they range between 5 and
10 km s$^{-1}$, in the southernly direction. 

The values of the internal ring velocity dispersions, $c$, that were
used to fit these solutions range between 1.0 and 1.3 km s$^{-1}$. 

The other
parameter that occurs in the wind solution is the ratio of the
wind luminosity $L$ to the ambient density, $\rho_0$. According to 
equation (21) in 
Weaver et al. (1977), which is appropriate 
for a thin shell, this ratio is given by 
\begin{equation}
{{L}\over{\rho_0}}=3.87{{R^5}\over{t^3}}.\label{eq:lrho}\end{equation}
Prior to the time of star formation in our model, this ratio was
$(2.2,4.2,1.6)\times10^{60}$ erg cm$^3$ s$^{-1}$ gm$^{-1}$ for the
spherical case in the post-collapse pressurized, snowplow, and constant
velocity solutions, respectively, and $(1.2,2.6,0.9)\times10^{60}$ erg
cm$^3$ s$^{-1}$ gm$^{-1}$ in the three corresponding cylindrical cases.
In units of $1.3\times10^{36}/(1.4m_H)$ erg cm$^3$ s$^{-1}$ gm$^{-1}$,
which is approximately the number of OB-star winds (Snow \& Morton
1976) per unit external hydrogen density, these ratios are $(4.0,7.5,2.9)$
and (2.2,4.7,1.8), respectively. Further multiplication by $n_H=1$
cm$^{-3}$ gives the effective number of OB-star winds, which, for our
assumed density and times, ranges between 2 and 8. These are typical  
numbers of stars with strong winds 
in OB associations, and not inconsistent with the
6 AI stars presently near the center of the Quadrant arc.

The mass of the gas cloud that made Quadrant 
can be estimated from the
ambient column density ($1.2\times10^{21}$ cm$^{-2}$), 
the triggering radius ($\sim200-280$ pc), and the section of the
circle that Quadrant represents ($1/4$). This mass is
$4.2-8.2\times10^5$ M$_\odot$.
The total cluster mass
in the arc is
$3.2\times10^5$ M$_\odot$,
based on data in Table 1 along with
Figure 13 in Girardi et al. (1995).
This implies that the overall efficiency for star formation
was around $\sim40-80$\% in this triggered region.
This seems high, but Quadrant is very dense with stars, and the numbers are 
imprecise. 
Also, the absence of HI gas close to Quadrant
(Kim et al. 1997) suggests a high efficiency.

\subsection{HI shell expansion after the Quadrant arc forms}

The HI ring currently in the vicinity of Constellation III is
larger than the radius of the Quadrant arc of stars. There is also
recent star formation along the perimeter of this ring, and there is
an unusually low density of HI in the ring center. In our model,
the HI void and the 
large size of the current HI ring are the result of continued
expansion of the gaseous structure
that originally made the Quadrant, driven in
more recent times by pressure from the Quadrant stars themselves.
Presumably these stars drove away the remaining 
gas that directly formed
Quadrant, and then continued to exert a pressure on the surrounding
medium, re-inflating the original cavity with hot gas from the
Quadrant's stellar winds and supernovae, and making the deep HI hole in
the center of the ring. Additional pressure from supernovae in the first
generation of stars would have been available too. 

We can model this second generation expansion using the original equations
for a wind-driven cavity (Pikelner 1968; Dyson 1973; Weaver et al. 1977), 
but with a solution modified to include an initial non-zero radius
and velocity. 
These are equations (17), (18), and (19) in Weaver et al.
(1977): 
\begin{eqnarray}
E=2\pi R^3P \\ {{dE}\over{dt}}=L-4\pi R^2PV \\
{{d}\over{dt}} \left( {{4\pi}\over{3}} R^3\rho_0 V\right)=4\pi R^2P,
\end{eqnarray}
for energy $E$, radius $R$, velocity $V$, 
pressure $P$, wind luminosity $L$, and ambient
density $\rho_0$.  These equations can be reduced to the single equation
\begin{equation}
{1\over 3}R^4{{d^2V}\over{dt^2}}+4R^3V{{dV}\over{dt}}+5R^2V^3=
{{L}\over{4\pi\rho_0}}.
\end{equation}
We take the initial conditions for these evolution equations to be the
state of the previous generation ring at the time of star formation 
in Quadrant, i.e., $V=V_{\rm trig}$, and $R=R_{\rm trig}$,
and we take the initial acceleration from the first generation
snowplow solution, considering that the cavity pressure
at the time of the Quadrant formation was much less 
than the new cavity pressure from the Quadrant stars. This gives
the initial condition $dV/dt=-3V_{\rm trig}^2/R_{\rm trig}$. 

\begin{figure}
\vspace {3.0in}
\includegraphics{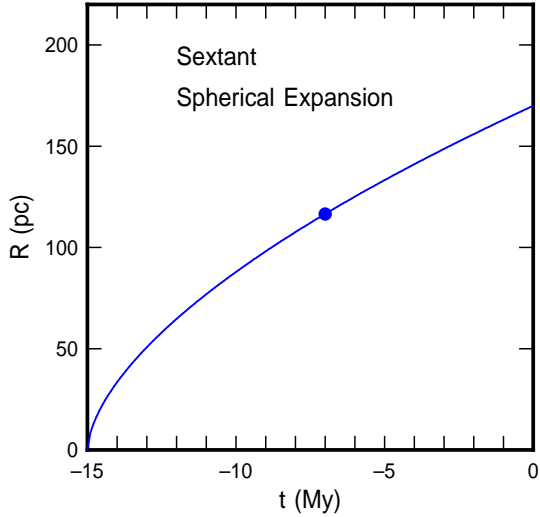}
\caption{The radius as a function of time for the proposed shell
that made the Sextant ring.
}
\end{figure}

The solution to this equation was determined numerically 
for the six cases considered above, and the resulting radii $R(t)$ 
for the re-inflated HI rings are shown in figure 3 as lines branching
off from the dots and going to 
larger radii. We adjust the only free parameter, $L/\rho_0$, to
give the current radius of the HI ring, which is 550 pc
(again assuming a distance to the LMC of 45 kpc). 
Written 
in terms of the first-generation ratio $L/\rho_0$ for the same $\rho_0$, 
the second-generation ratio is larger in these solutions by the factors
(156,78,220) and (286,130,368), for the spherical and cylindrical
cases, respectively, in the 
pressurized, snowplow, and constant
velocity solutions.  Thus the Quadrant arc is more luminous than the
first generation OB association by these factors of $\sim80$ to $\sim370$,
considering that the average density outside the HI ring was about
the same in each case.
This large factor, combined with the younger age of Quadrant, explains why
the pressure-driving cluster is so difficult to see in comparison to 
the Quadrant arc. 

The current expansion speed of the HI shell in these solutions 
is $\sim19$ km s$^{-1}$, tightly constrained by the
current radius and assumed age of the ring, and by the radius and age of 
the ring when Quadrant formed. The projected expansion speed of this HI
ring is predicted to be about half of 
this, considering the LMC inclination (33$^\circ$). This is a small
enough expansion speed to be consistent with the observations
(Domg\"orgen et al. 1995).

\section{Triggered star formation in Sextant}

The main event in this region of the LMC is the expansion of the giant
HI ring that made Quadrant, and the continued expansion of this
ring afterwards.  
Many other star formation sites have also appeared in the ring,
particularly at 
later stages when the ring is dense and strongly self-gravitating. 
One of these may have gone on to trigger another arc of
stars in the southwest, which we call Sextant. 
This second arc is small enough that the
spherical solution for an expanding shell is probably adequate,
in which case we can use the results for gravitational collapse
directly from equations (16) and (17) in Elmegreen (1994), which are
\begin{equation}
t_{\rm trig}={{1.25}\over{\left(G\rho_0 {\cal M}\right)^{1/2}}} \;\;;\;\; 
R_{\rm trig}={5\over 3}Vt_{\rm trig}=
{{2.1c{\cal M}^{1/2}}\over{G\rho_0}} .
\label{eq:shell}
\end{equation}

A plausible model for the formation of Sextant is that it was triggered
by the collapse of a small swept-up shell 
around a cluster that formed
inside the giant expanding HI ring. Then Sextant is a third generation
of star formation, and the driving cluster that made Sextant is a
second generation, like Quadrant, but younger. 

The oldest stars in Sextant are $\sim7$ My old, according to table 1, and
the clusters near the center of the Sextant arc, either HS287 or HS 288, 
are probably only 10-15 My old, considering HS 287 still has an HII region. 
Thus Sextant could have been triggered 
in only $\sim8$ My ($=15$ My $-7$ My).  This means, according to equation 
(\ref{eq:shell}), that the ambient density had to be rather large.
This is to be expected if Sextant was triggered inside the 
dense part of the HI ring. 

We can find the ratio of the 
external density to the internal sound speed, 
$n_H/c$, for the shell that made Sextant using equations
(\ref{eq:shell}) and (\ref{eq:rnow1}), with the constraints that
the triggering occurred $t_{\rm trig}=8$ My after the 
first generation formed, which was
$t_{\rm now}=15$ My ago, and that the current
size of the shell is $R_{\rm now}=170$ pc. These numbers are not
well known, but they are good enough
to illustrate the procedure. Then combining these equations to obtain 
$V_{\rm trig}$,
we first get $t_{\rm trig}^{2/5}=(3/5)R_{\rm now}t_{\rm now}^{-3/5}
V_{\rm trig}^{-1}$ by eliminating $R_{\rm trig}$ from
equation (\ref{eq:rnow1}) and the right hand equation (\ref{eq:shell}),
and then, by substituting $t_{\rm trig}$ from the 
left hand equation (\ref{eq:shell}), we get
\begin{equation}
V_{\rm trig}=\left({{3R_{\rm now}}\over{5t_{\rm now}^{3/5}}}\right)^{5/4}
\left({{(G\rho_0)^{1/2}}\over{1.25c^{1/2}}}\right)^{1/2}.
\end{equation}
Putting this into the 
${\cal M}$ term of the 
left hand equation (\ref{eq:shell}),
we get
\begin{equation}
t_{\rm trig}={{1.25^{5/4}c^{5/8}}\over{(G\rho_0)^{5/8}}}
\left({{5t_{\rm now}^{3/5}}\over{3R_{\rm now}}}\right)^{5/8}.
\end{equation}
For $c$ in km s$^{-1}$, $R_{\rm now}$ in pc, 
$n_{\rm H}$ in cm$^{-3}$, and $t_{\rm now}$ in 
My, this is
\begin{equation}
t_{\rm trig}=443 \left({c\over{n_H}}\right)^{5/8}
\left({{t_{\rm now}^{3/5}}\over{R_{\rm now}}}\right)^{5/8} \;\;{\rm My}.
\end{equation}
Now we set $t_{\rm trig}=8$ My, $t_{\rm now}=15$ My, and
$R_{\rm now}=170$ pc, to get $n_H/c\sim18$ cm$^{-3}$(km s$^{-1}$)$^{-1}$.
This large value indicates how the density in the environment of Sextant
was likely to be large for $c\sim1$ km s$^{-1}$. 

The triggering radius follows from these equations in a similar manner:
\begin{equation}
R_{\rm trig}={5\over 3}\left({{1.25c^{1/2}}\over{(G\rho_0)^{1/2}}}
\right)^{3/4}\left({{3R_{\rm now}}\over{5t_{\rm now}^{3/5}}}
\right)^{5/8},
\end{equation}
which may be re-written in units of km s$^{-1}$, pc, and My:
\begin{equation}
R_{\rm trig}=38.7\left({{c}\over{n_H}}\right)^{3/8}
\left({{R_{\rm now}}\over{t_{\rm now}^{3/5}}}\right)^{5/8}\;\;{\rm pc}.
\end{equation}
With $t_{\rm now}=15$ My, $R_{\rm now}=170$ pc, and
$n_H/c\sim18$ cm$^{-3}$(km s$^{-1}$)$^{-1}$, this gives 
R$_{\rm trig}=117$ pc.  

Finally, the velocity at the time of triggering becomes simply
$V_{\rm trig}=(3/5)R_{\rm trig}/t_{\rm trig}$, which is
$\sim8.7$ km s$^{-1}$. 

Figure 4 shows the solution $R(t)$ with the time of triggering for
Sextant, using $n_H/c\sim18$ cm$^{-3}$(km s$^{-1}$)$^{-1}$. The
corresponding ratio $L/\rho_0$ for this shell solution is
$1.4\times10^{60}$ erg cm$^3$ s$^{-1}$ gm$^{-1}$ from equation
\ref{eq:lrho}, corresponding to a number of OB stars equal to $\sim2.6$
per unit external HI density, or to $\sim48$ stars if $n_H\sim18$
cm$^{-3}$. If this is too large for the clusters HS287 and HS 288, then 
perhaps these clusters are slightly older than we assumed (making the
power requirement smaller), or there was significant supernova activity
in the center of the Sextant arc, in addition to stellar winds. 

\section{Low shear and large disk thickness as a pre-requisite for forming
giant shells and rings}

The large size and round shape of the LMC4 superbubble is most likely
the result of low shear in this region. Otherwise, the ring resembles
the Lindblad ring in the Solar neighborhood in overall dimension and
mass, even to the extend that the Lindblad ring also has an old and
faint dispersed OB association in the center (Blaauw 1984) with
significant star formation along the periphery (P\"oppel 1997). 
The low shear in the LMC makes the LMC4 region morphologically
different than the Lindblad ring, however. 

The rotation speed in the LMC at the radius of LMC4, $R\sim2.6$ kpc, is
about $V\sim 55$ km s$^{-1}$ deprojected, and it is in a flat part of the
rotation curve (Luks \& Rohlfs 1992). Thus the shear time, which is the
inverse of the Oort A parameter, is $A^{-1}\sim2R/V\sim94$ My. This is
$\sim3$ times larger than the age of the HI ring according to our model
($\sim30$ My), and so the ring is still nearly circular. Generally, the
pitch angle $i$ of an initially circular feature that shears with time
$t$ is given by $\tan i = -tRd\Omega/dR$ for angular rate $\Omega=V/R$.
For the LMC4 region after $t=30$ My, this pitch angle is 58$^\circ$,
which is sufficiently close to the radial direction (90$^\circ$) that
the ring hardly looks swept back, especially with the $\sim33^\circ$
inclination of the galaxy. 

Shear is generally low in other dwarf galaxies too, such as HoII, where
other giant rings have been found (Puche et al. 1992). These rings can
be many tens of millions of years old and still nearly circular. At such
large ages, the central clusters may be dispersed over regions
several hundred parsecs in diameter. The only obvious tracers of these
first generation stars would
be supergiants and Cepheids, as in the central
region of LMC4. The stars may even disperse from their clusters faster
than the average speed of the ring at very late times, after the ring
has stalled (e.g., $>50$ My). Then the brightest stars from all of the
neighboring expansion regions can mix together, smoothing out the
initial associations. The main sequence stars that formed in these
associations will blend with the older stars in the disk.

Kiloparsec-size rings can also occur in giant galaxies, but primarily in
the outer spiral arms, where the shear and flow-through times are very
large. Shear is generally low inside spiral arms because of angular
momentum conservation in the gas during the spiral wave compression
(Elmegreen 1992). In the outer regions, particularly near corotation or
beyond, the flow-through time can exceed 100 My. Then star forming
regions can inflate giant shells for several generations. Examples of
this might be the giant shells in the southern spiral arm of M83
(Sandage \& Bedke 1988), and in the northern spiral arm of M51 (see the
15$\mu$-B image in Block et al. 1997).

A second condition for the formation of giant shells and rings is that
the disk thickness has to exceed the perturbation diameter of the high
pressure region, or else the high pressure gas will escape into the halo
(MacLow \& McCray 1988; Tenorio-Tagle, Rozyczka, \& Bodenheimer 1990). 
Such large disk thicknesses are generally believed to be
appropriate for dwarf galaxies like the LMC 
(Hodge \& Hitchcock 1966; van den Bergh 1988; Puche et at. 1992) 
and for the outer
regions of giant spiral galaxies, because of the low surface
brightnesses, and therefore, low stellar surface densities, of the
underlying disks. With a low surface mass density in a disk, the $\sim5$
km s$^{-1}$ turbulent motions of the gas bring it to a large scale
height, which scales inversely with the total surface density inside the
gas layer. Rings can become even larger than the disk thickness after
the interior pressure decreases, simply by momentum conservation from
their motion in the plane. For example, if the shell speed at the time
of breakout is twice the external velocity dispersion, then the final
ring diameter will be $2^{1/2}$ times the disk thickness by the time the
expansion has slowed to the external dispersion; i.e., it will have
picked up twice the mass in that final stage, and therefore slow to half
the speed. 

In view of the low shear and large likely disk thickness in the outer part of
the LMC, it is not surprising that several supergiant rings occur there,
and that these rings today have only faint remnants of the powerful
stars that once created them. 

\section{Conclusions}

Two arcs of star clusters inside and on the rim of the
superbubble LMC 4, including what is commonly referred to as
Constellation III,  
may have been formed by the self-gravitational
collapse of gas in swept-up pieces of rings or shells. The dimensions
and time sequences for these triggerings are reasonable, as is the
energetics. This triggering model differs qualitatively from others in
which pre-existing clouds are squeezed into star formation by passing
shock fronts. The regular form of the Quadrant and Sextant arcs
resembles more a piece of a ring or shell than a random arrangement of
pre-existing clouds. 

The conditions in the outer part of the LMC and other dwarf galaxies,
and in the outer spiral arms of giant spiral galaxies, are favorable for
the formation of giant gas shells and rings in which the first
generation of stars is so old and dispersed that it is barely visible
anymore. All that is required for this is a low rate of shear and a
relatively large disk thickness. 

A recent study suggests that Gamma Ray Bursts from old star-forming
reigons might also play a role in enlarging supernova cavities 
beyond the size of the disk thickness (Efremov,
Elmegreen \& Hodge 1998). This would not affect the general
triggering scenario discussed here, but it would affect the estimate
for the number of stars that led to either the first or the
second generation pressures. 

We are grateful to the referee for valuable suggestions
and to P.Bogdanovsky, whose computer program helped in searches for
triggering stars. Dr. K. Olsen kindly provided a tiff image
for the star field in figure 1.

\begin{table*}
 \centering
 \begin{minipage}{140mm}
  \caption{Clusters within arcs}
  \begin{tabular}{@{}llrrrrrrrr@{}}  

Arc &Cluster &V &U-B &B-V &X ($^\circ$) &Y ($^\circ$) &S &$\log t$\\
Quadrant&&&&&&&&&\\
&KMK987 	&12.32 &-0.73 &-0.12 &-0.91 &2.59     &15 &7.33\\
&NGC2002=SL517	&10.10 &-0.58 &0.34  &-0.89 &2.61     &13 &7.18\\
&SL538		&11.30 &-0.67 &-0.01 &-0.98 &2.53     &15 &7.33\\
&NGC2006=SL537	&10.88 &-0.63 &0.12  &-0.98 &2.52     &15 &7.33\\
&KMK1019	&11.49 &-0.57 &0.48  &-1.03 &2.49     &12 &7.11\\
&LH77p1 	&10.55 &-0.64 &0.19  &-0.97 &2.47     &13 &7.18\\
&LH77p2 	&10.09 &-0.65 &0.10  &-1.26 &2.49     &14 &7.25\\
&LH77p3 	&10.39 &-0.73 &0.04  &-1.09 &2.51     &13 &7.18\\
&NGC2027=SL592	&10.97 &-0.89 &0.000 &-1.34 &2.58     & 8 &6.82\\
&KMK1074	&12.61 &-0.80 &-0.15 &-1.24 &2.49     &13 &7.18\\
&SL586=ESO86sc12&11.17 &-0.80 &-0.12 &-1.31 &2.53     &13 &7.18\\
&NGC2034n	& 9.78 &-0.58 &0.29  &-1.41 &2.62     &14 &7.25\\
&NGC2034s	&10.35 &-0.75 &0.24  &-1.39 &2.58     & 9 &6.89\\

Sextant&&&&&&&&&\\
&SL456=LH51	&11.75 &-1.02 &-0.23  &-0.41 &2.01     &7 &6.74\\
&NGC1955=LH54	& 9.83 &-1.00 &-0.21  &-0.46 &1.99     &7 &6.74\\
&NGC1968w=LH60w &10.78 &-0.96 &-0.04  &-0.62 &2.03     &5 &6.60\\
&NGC1968e=LH60e &10.20 &-1.06 &-0.21  &-0.60 &2.03     &4 &6.52\\
&NGC1974=LH63	&10.30 &-0.97 &-0.21  &-0.65 &2.06     &8 &6.82\\
Center&HS288	&      &      &       &-0.55 &2.25     &  &    \\
      &HS287	&      &      &       &-0.48 &2.35     &  &    \\

Third Arc &&&&&&&&&\\

&Inside N59:&&&&&&&&\\
&NGC2040 in LH82 &11.47  &-0.95  &-0.18 &-1.42 &1.93  &9  &6.89 \\
&NGC2035 in LH82 &10.99  &-0.76  &-0.17 &-1.37 &1.91  &15 &7.33 \\
&NGC2032 in LH82 &10.80  &-0.58  &-0.19 &-1.35 &1.93  &19 &7.62 \\
&NGC2029 in LH82 &12.29  &-0.68  &-0.39 &-1.32 &1.94  &19 &7.62 \\

&Inside N56:&&&&&&&&\\
&NGC2021 in LH79 &12.06  &-0.77  &-0.13 &-1.18 &2.04  &14 &7.25 \\
&SL567  in LH78  &10.19  &-0.83  &0.09  &-1.15 &1.97  &9  &6.89 \\
&NGC2011 in LH75 &10.58  &-0.71  &0.04  &-1.06 &1.98  &13 &7.18 \\

&NGC2004         &9.60  &-0.71  &0.13   &-0.91  &2.20  &12  &7.11 \\ 
&SL522           &12.10 &-0.78  &-0.09  &-0.91  &2.30  &13  &7.18 \\
&SL516           &12.14 &-0.61  &-0.05  &-0.85  &2.51  &17  &7.47 \\
&NGC2002=SL517 	 &10.10 &-0.58  &0.34   &-0.89  &2.61  &13  &7.18 \\
Center&NGC2041   &10.36  &-0.17   &0.22  &-1.48  &2.51  &24 &7.99 
\end{tabular}
\end{minipage}
\end{table*}

\begin{table*}
 \centering
 \begin{minipage}{140mm}
  \caption{A-type Supergiant stars near the center of the LMC4 supershell}
  \begin{tabular}{@{}lcccc@{}}  
Star &R.A. (2000.0) &Dec (2000.0) &V &Sp.T.\\
NS 119A-66 & 5$^{\rm h}$ 31${\rm m}$    &-66$^\circ$ 46'  &12.16  &A0 I  \\
NS 120 -66 & 5$^{\rm h}$ 31${\rm m}$    &-66$^\circ$ 41'  &12.50  &A0 I  \\  
NS 124 -66 & 5$^{\rm h}$ 31.5${\rm m}$  &-66$^\circ$ 42'  &12.46  &A1 I  \\ 
G  359     & 5$^{\rm h}$ 32.1${\rm m}$  &-66$^\circ$ 39'  &12.51  &A9 I  \\  
NS 129 -66 & 5$^{\rm h}$ 32.2${\rm m}$  &-66$^\circ$ 43'  &12.16  &A0 I  \\  
NS 130 -66 & 5$^{\rm h}$ 32.2${\rm m}$  &-66$^\circ$ 40'  &12.05  &A0 I  
\end{tabular}
\end{minipage}
\end{table*}

\bsp

\label{lastpage}

\end{document}